**Graphical Abstract**

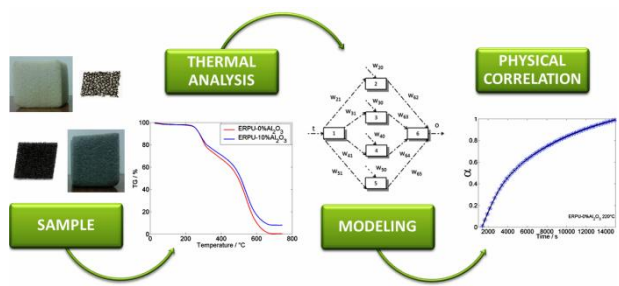

Thermal analysis data of two Rigid Polyurethane Foam samples: loaded with ($Al_2O_3$) and no inorganic filler are modeled by neural network supporting physical correlation.

# ESTUDO CINÉTICO DE DECOMPOSIÇÃO TÉRMICA DE ESPUMAS RÍGIDAS DE POLIURETANO POR REDE NEURAL ARTIFICIAL

**Bárbara D. L. Ferreira[a], Virgínia R. Silva[b], Maria Irene Yoshida[a], Rita C.O. Sebastiao [a,]\***

[a]Departamento de Química, Universidade Federal de Minas Gerais - UFMG, 31270-901 Belo Horizonte - MG,Brasil

[b]Centro de Desenvolvimento da Tecnologia Nuclear - CDTN/CNEN, 31270-901 Belo Horizonte – MG, Brasil

\*e-mail: ritacos@ufmg.br

# KINETIC STUDY OF RIGID POLYURETHANE FOAMS THERMAL DECOMPOSITION BY ARTIFICIAL NEURAL NETWORK

Kinetic models of solid thermal decomposition are traditionally used for individual fit of isothermal decomposition experimental data. However, this methodology can provide unacceptable errors in some cases. To solve this problem, a neural network (MLP) was developed and adopted in this work. The implemented algorithm uses the rate constants as predetermined weights between the input and intermediate layer and kinetic models as activation functions of neurons in the hidden layer. The contribution of each model in the overall fit of the experimental data is calculated as the weights between the intermediate and output layer. In this way, the phenomenon is better described as a sum of kinetic processes. Two rigid polyurethane foam samples: loaded with $Al_2O_3$ and no inorganic filler were used in this work. The $R_3$ model described the thermal decomposition kinetic process for all temperatures for both foams with smaller residual error. However, the network residual errors are on average $10^2$ times lower compared to this individual kinetic model. This improved methodology allows detailed study of physical processes and therefore a more accurate determination of kinetic parameters such as the activation energy and frequency factor.

**INTRODUÇÃO**

Estudos cinéticos de decomposição térmica no estado sólido são de grande interesse científico e industrial.[1] Três metodologias relevantes no tratamento teórico de cinética de decomposição em sólidos são: (i) descrição do processo por modelos cinéticos, (ii) associação e correção dos modelos cinéticos por rede neural artificial e (iii) modelo isoconversional.

A metodologia de ajuste por modelos cinéticos consideram que as reações de decomposição térmica de sólidos ocorrem na interface do produto-reagente. Este processo cinético, com base na formação e crescimento de núcleos, pode ser estudado por análise de dados de decomposição térmica nos quais a redução da massa é medida num intervalo de tempo à temperatura constante. Os núcleos de reação são preferencialmente formados em imperfeições da estrutura e os modelos cinéticos são usados para explicar as curvas experimentais. Um determinado modelo é escolhido devido a sua melhor correlação para ajustar os dados experimentais. No entanto, em muitas situações, o erro residual do modelo não é aceitável na descrição do processo total, embora, para regiões específicas de decomposição, o modelo possa ser apropriado, sendo apenas uma primeira aproximação. Os modelos podem ser representados por uma equação geral do tipo:

$$\frac{d(1-\alpha)}{dt} = -k(1-\alpha)^n \big(1 - q(1-\alpha)\big)^m \tag{1}$$

Em que $\alpha$ representa a fração de decomposição, $t$ o tempo, $k$ a constante de velocidade e $q$ é o parâmetro de iniciação. Os parâmetros de ajuste $n$ e $m$ definem o modelo físico.

Em alguns trabalhos já publicados pelo nosso grupo, foi construído um programa computacional baseado na teoria de redes neurais de múltiplas camadas para determinar o conjunto de modelos que melhor descrevem a cinética de decomposição térmica de sólidos. Nestes trabalhos foram estudados a cinética de decomposição do acetato de Ródio e os fármacos Lamivudina e Efavirenz, usados no coquetel anti-HIV.[2-4] Esta análise permite a correção matemática do modelo cinético e a determinação do processo físico global com erro experimental de ajuste das curvas bem menor que a utilização de apenas um dos modelos, o que consequentemente fornece informações preciosas sobre o material estudado. O método consiste na linearização da rede neural pela fixação dos pesos de interconexão entre a camada de entrada e a camada intermediária, fazendo com que cada neurônio da camada oculta represente um modelo cinético a ser ponderado pelo neurônio na camada de saída. A rede neural representa um poderoso método de rotina para estudar este tipo de reação.

Outra abordagem, descrita por Friedman, consiste em tratar os materiais por técnicas termogravimétricas em que a fração de decomposição é medida em função da temperatura á diferentes razões de aquecimento do sólido. Curvas experimentais são comparadas e os valores de decomposição, comuns a todas as curvas, são tratados a fim de se obter os parâmetros cinéticos (energia de ativação e fator de frequência) do processo. A equação geral que descreve esta metodologia é

$$\ln(-\frac{d(1-\alpha)}{dt}) = -\frac{E}{RT} + \ln(A(1-\alpha)^n) \qquad (2)$$

No presente trabalho estaremos interessados em investigar a cinética de decomposição de espumas rígidas de poliuretano (ERPU), sem e com adição de carga inorgânica contendo grande quantidade de $Al_2O_3$, pela metodologia de redes neurais. As espumas rígidas de poliuretano (ERPU) é uma das diversas formas em que se encontra o poliuretano (PU), um polímero formado essencialmente pela reação de condensação de um isocianato com um poliol. Nesta reação há dispersão de um gás, causada pelo aquecimento e evaporação de um agente de expansão durante o processo de polimerização, que origina uma estrutura tridimensional altamente reticulada de baixa densidade. As ERPU podem ser usadas em ampla faixa de temperatura, de -200°C a +150°C, por possuir alto ponto de amolecimento, além de ser resistente a produtos químicos e apresentar um baixo coeficiente de condutividade térmica, em virtude do gás de expansão preso nas unidades celulares e da baixa condutividade do próprio polímero. Devido a estas propriedades, é um dos mais eficientes isolantes térmicos usados na construção civil. Sua desvantagem é ser altamente inflamável, mas esta característica pode ser contornada com a adição de substâncias que retardam a propagação das chamas, como dibromo propanol, o tris(dicloropropilfostato), (TCPP) e a alumina.[5-11,24]

Em geral, os aditivos retardantes de chama são substâncias químicas que ao serem adicionadas aos polímeros interferirão nas condições da combustão. Essas substâncias podem ser classificadas de acordo com sua composição química como retardantes orgânicos reativos ou não reativos e os inorgânicos. Os dois primeiros apresentam compostos de fósforo e de halogênios, em especial cloro e bromo, e se diferenciam apenas pela presença ou não desses grupos funcionais presos quimicamente à cadeia polimérica. Dentre os retardantes de chamas inorgânicos podem ser citados o hidróxido de alumínio ou alumina trihidratata, material inerte de alta estabilidade. A alumina é preferível frente aos compostos contendo halogênio e fósforo que são extremamente tóxicos e prejudiciais ao meio ambiente.[11-17]

# TEORIA DOS MODELOS CINÉTICOS

A cinética da decomposição térmica de materiais é baseada na nucleação e no crescimento dos núcleos ativos presentes na superfície do cristal.[18] A existência de pontos reativos separados, relacionados com as imperfeições dos cristais, provoca um aumento da energia de Gibbs do sistema e, consequentemente, a sua reatividade.[19] Diversos modelos físicos são utilizados para descrever a fração decomposta, $\alpha$, em função do tempo de decomposição. Sendo $\alpha$ definido como a quantidade de perda de massa no tempo $t$ normalizado para a massa total perdida.[20] Os modelos físicos estão apresentados na Tabela 1 e em geral são classificados de acordo com a forma da curva e com o tempo de aceleração e desaceleração.[19,21]

**Tabela 1.** Modelos Cinéticos de Decomposição Térmica

| Modelos | Símbolos | Equações Cinéticas |
|---|---|---|
| Aceleração | | |
| Lei Potencial | $P_n$ | $\alpha^{1/n} = kt + k_0 \quad n = 2,3,4,\ldots$ |
| Sigmoide | | |
| Avrami-Erofeev | $A_m$ | $[-ln(1-\alpha)]^{1/m} = kt + k_0$ <br> $m = 2, 3, 4\ldots$ |
| Avrami-Erofeev | $A_u$ | $\ln \dfrac{\alpha}{1-\alpha} = kt + k_0$ |
| Prout-Tompkins | $A_x$ | $\ln \dfrac{\alpha}{1-\alpha} = k \ln t + k_0 \quad k > 1$ |
| Desaceleração | | |
| Modelo geométrico - contração | | |
| Contração Linear | $R_1$ | $1-(1-\alpha) = kt + k_0$ |
| Contração Superficial | $R_2$ | $1-(1-\alpha)^{1/2} = kt + k_0$ |
| Contração Volumétrica | $R_3$ | $1-(1-\alpha)^{1/3} = kt + k_0$ |
| Modelo de Difusão | | |
| Uma Dimensão | $D_1$ | $\alpha^2 = kt + k_0$ |
| Duas Dimensões | $D_2$ | $(1-\alpha)\ln(1-\alpha) + \alpha = kt + k_0$ |

| | | | |
|---|---|---|---|
| Três Dimensões | $D_3$ | $\left[1-(1-\alpha)^{1/3}\right]^2 = kt+k_0$ | |
| Ginstling-Brounshtein | $D_4$ | $1-\dfrac{2\alpha}{3}-(1-\alpha)^{2/3} = kt+k_0$ | |

As reações de decomposição, em que predomina a etapa de aceleração, se caracterizam por apresentar apenas o fenômeno de nucleação. Nessas reações a formação de núcleos pode ocorrer de forma instantânea ou com velocidade de nucleação constante e o modelo exponencial é mais usado para descrever o evento.[18,21] Já as reações em que se observa uma nucleação caótica seguida do crescimento desses núcleos, os modelos de Avrami-Erofeev ou Prout-Tompkins são mais utilizados.[22]

Nas reações onde há apenas o crescimento dos núcleos, as curvas de desaceleração são mais apropriadas e dois fenômenos são responsáveis pela cinética dessas decomposições: a contração e a difusão. A contração é responsável pelo desenvolvimento rápido de núcleo em toda a extensão da superfície do cristal. Já a difusão é responsável pelo controle da taxa de reação, uma vez que a continuidade da reação requer o transporte dos reagentes para a camada de produto.[18-22]

A determinação do modelo físico no processo de decomposição de materiais é crucial para o estudo da cinética das reações e pode ser feita por análise de microscopia, através de softwares comerciais, geralmente restritos a no máximo três modelos para ajustar os dados experimentais ou por redes neurais artificiais, que ponderam a contribuição de todos os onze modelos físicos no processo.[2-4,23]

## REDE NEURAL MLP APLICADA A DECOMPOSIÇÃO TÉRMICA DE SÓLIDOS

A metodologia adotada nesse trabalho utiliza um algoritmo proposto originalmente por Sebastião e colaboradores em 2003 e em 2004 para o estudo da decomposição térmica do acetato de ródio e acetado de ródio(II), respectivamente. A rede MLP possui três camadas, uma de entrada, uma intermediária e uma de saída. Nas camadas de entrada e de saída há apenas um neurônio e na camada intermediária o número de neurônios é variável, dependendo da quantidade de modelos cinéticos que se deseja incluir no estudo. A metodologia possui três etapas principais: a primeira etapa do algoritmo consiste em fixar os pesos de interconexão entre os neurônios da camada de entrada e a camada intermediária e os bias dos neurônios na camada intermediária.[2-4] Essas constantes são armazenadas no vetor **W₁**, responsável pela linearização da rede,

$$\mathbf{W_1 X} = \begin{pmatrix} w_{21} + w_{20} \\ w_{31} + w_{30} \\ w_{41} + w_{40} \\ w_{51} + w_{50} \end{pmatrix} \quad (3)$$

Os valores $w_{i1}$ e $w_{i0}$ são os pesos de interconexão entre a camada de entrada e a camada intermediária e correspondem aos valores das constantes $k$ e $k_0$, nessa ordem, obtidos pelos modelos cinéticos no ajuste experimental dos dados. Essa etapa é importante para que não ocorra perda de informação química ao longo do processo, além de garantir que os modelos mostrados na Tabela 1 possam ser usados como função de ativação para cada neurônio da camada intermediária dessa rede.

Já na segunda etapa do algoritmo, é realizada uma transformação não-linear em que cada neurônio da camada intermediária é ativado por uma função de ativação, *f*, correspondente ao modelo cinético. Como resultado tem-se o vetor **B** que determina os estados dos neurônios ativados como mostra a Equação 4.[2-4]

$$\mathbf{B} = f\,(\mathbf{W_1 i}) \quad (4)$$

Nesta notação, **i**, representa o estado do neurônio da camada de entrada, ou seja os dados experimentais de *t*. As funções matemáticas utilizadas para ativar os neurônios são os modelos cinéticos descritos na Tabela 1. Essas funções matemáticas são adequadas por terem similaridade com o impulso nervoso. Dessa forma, os estados de neurônios intermediários são representados na Equação 5. Vale ressaltar que essa estrutura de rede não é rígida, sendo possível adicionar ou excluir alguns modelos da camada intermediária, dependendo do sistema em estudo.[4]

$$\mathbf{W_1 x} = \begin{pmatrix} D_1(w_{21}t + w_{20}) \\ D_2(w_{31}t + w_{30}) \\ D_3(w_{41}t + w_{40}) \\ D_4(w_{51}t + w_{50}) \\ R_1(w_{61}t + w_{60}) \\ R_2(w_{71}t + w_{70}) \\ R_3(w_{81}t + w_{80}) \\ Am_4(w_{91}t + w_{90}) \\ Am_2(w_{101}t + w_{100}) \\ AU(w_{111}t + w_{110}) \\ F_1(w_{121}t + w_{120}) \end{pmatrix} \quad (5)$$

Cada neurônio intermediário, $i_2$ até $i_{12}$, tem sua contribuição para o estado do neurônio da camada de saída, $i_{13}$, sendo ponderados pelos pesos de interconexão entre a camada intermediária e a de saída pelo vetor $\mathbf{W_2}$. O vetor $\mathbf{W_2}$ pode ser exemplificado como apresenta a Equação 6.

$$\mathbf{W_2} = (w_{132}\ w_{133}\ w_{134}\ w_{135}\ ....\ w_{1312}) \tag{6}$$

E na terceira e última etapa ocorre uma transformação linear, resultando no vetor $\mathbf{Y}$ que é a resposta da rede,

$$\mathbf{Y} = \mathbf{W_2} f(\mathbf{W_1 i}) \tag{7}$$

Assim, a função erro da rede pode ser calculada pela diferença entre o estado do neurônio da camada de saída e os dados de decomposição obtidos experimentalmente,

$$E = \left\| \mathbf{Y}_{cal} - \mathbf{Y}_{exp} \right\|_2^2 \tag{8}$$

Com essa estrutura de rede é possível calcular a contribuição dos diversos modelos cinéticos através do algoritmo de optimização por Levenberg-Marquart,

$$\mathbf{W_2} = (\mathbf{B^T B})^{-1} \mathbf{B^T Y}_{cal} \tag{9}$$

**PARTE EXPERIMENTAL**

As amostras de espumas rígidas de poliuretano (ERPU) utilizadas nesse trabalho contém glicerol como agente reticulador. A primeira amostra é uma ERPU sem carga (ERPU-0%$Al_2O_3$) e a segunda amostra recebeu cerca de 10% de carga inorgânica contendo alumina (ERPU-10%$Al_2O_3$). A carga utilizada é proveniente do filtro eletrostático da etapa de calcinação do processo Bayer, fornecido por indústrias de produção de alumínio do Estado de Minas Gerais.[24]

Ambas as espumas foram sintetizadas utilizando-se poliol de origem vegetal, produzido através da modificação química (esterificação) de óleo extraído das sementes de *Ricinus communis*, popularmente conhecida no Brasil como Mamona.[25,26] Essas espumas foram submetidas à decomposição térmica com intuito de obter suas curvas TG dinâmicas, mostradas na Figura 1. As curvas TG foram obtidas em uma termobalança TGA-50H SHIMADZU, em cadinho de alumina, atmosfera de ar síntético com fluxo de 100 mL/min e razão de aquecimento de 10 °C/min.

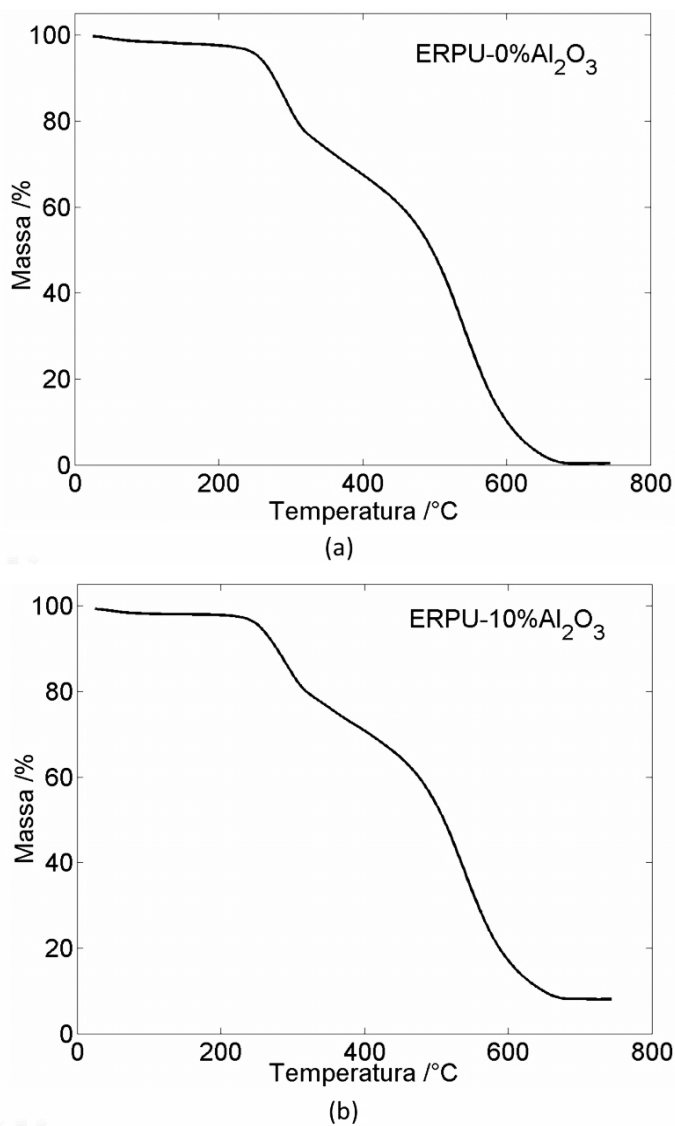

***Figura 1.*** *Curva TG dinâmica para amostras das espumas rígidas de poliuretano contendo glicerol a) sem adição de carga e b) com adição de 10% de rejeito industrial contendo alumina, respectivamente*

As curvas TG para as amostras das espumas rígidas de poliuretano contendo glicerol sem adição de carga e com adição de 10% de carga com alumina proveniente de rejeito industrial apresentadas na Figura 1 exibem perfis semelhantes, com dois estágios em seus processos de decomposição. A primeira etapa de perda de massa é relativa à degradação dos grupos uretanos que ocorre na faixa de 215°C a 350°C. Já o segundo estágio refere-se à degradação dos resíduos da primeira decomposição que ocorre na faixa de 400°C a 620°C para ambas as espumas.

Assim, como o processo de decomposição dos grupos uretanos ocorre em temperatura mais baixa, em torno de 228°C, essa foi escolhida por ser a primeira observação experimental, além de apresentar perda de massa significativa nessa temperatura, para as respectivas amostras.

A temperatura em que o processo de decomposição se inicia, para os grupos uretano, nas espumas sem e com adição de alumina é a mesma, uma vez que a adição de 10% carga não altera a estabilidade térmica dos compósitos, já que o resíduo inorgânico não altera quimicamente a matriz polimérica e a densidade de ligações cruzadas no polímero. Porém, testes de chama realizados nestas mesmas amostras mostram uma redução de cerca de 30% nos tempos de auto extinção da chama para as espumas com 10% de carga, quando comparadas às ERPU sem carga, o que evidencia o efeito retardante de chamas desta carga na matriz polimérica.

Quatro curvas TG isotérmicas experimentais, referentes ao início do processo de degradação dos grupos uretano, para cada ERPU, foram obtidas nas mesmas condições das curvas TG dinâmicas e são apresentadas na Figura 2.

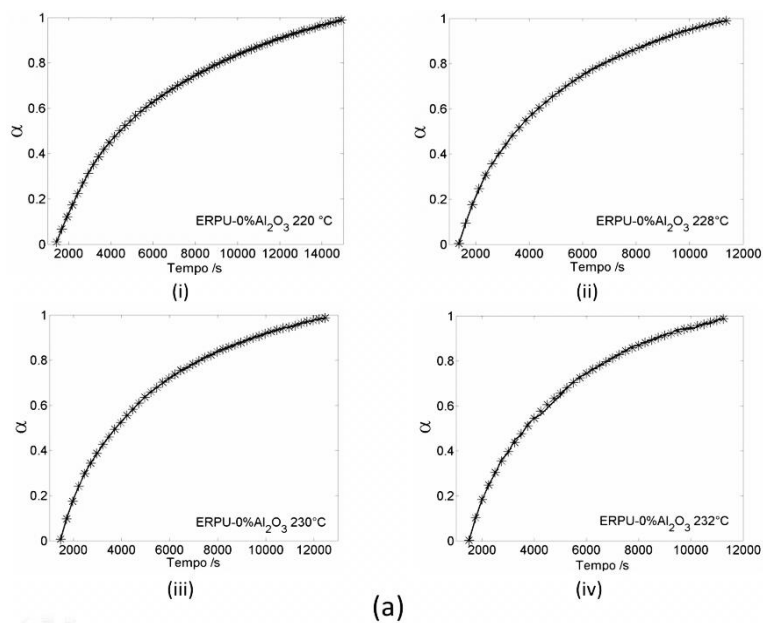

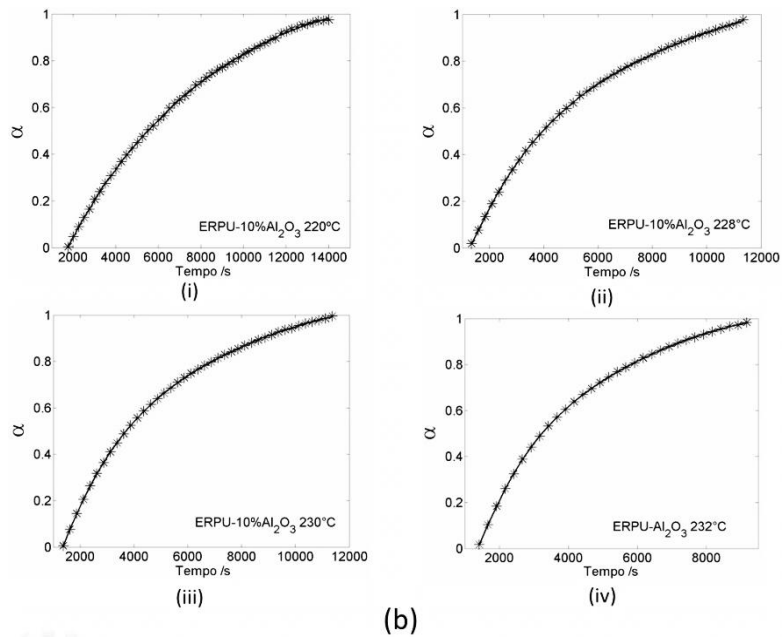

***Figura 2.*** *Dados experimentais (\*) e o ajuste fornecido pela rede (-) para as amostras de espumas rígidas de poliuretano contendo glicerol como agente reticulador, a) sem adição de carga contendo alumina (ERPU-0%Al$_2$O$_3$) e b) com adição de 10% de carga contendo alumina (ERPU-10%Al$_2$O$_3$) nas temperaturas: i)220°C, ii)228°C, iii)230° e iv)232°C, respectivamente*

As curvas isotérmicas experimentais, mostradas na Figura 2 nas temperaturas de 220°C, 228°C, 230° e 232°C para as amostras de espumas rígidas de poliuretano contendo glicerol como agente reticulador sem e com adição de 10% de carga contendo alumina, proveniente de rejeito industrial, foram usadas nesse trabalho para investigar a cinética de decomposição dos grupos uretanos presentes no polímero.

## RESULTADOS E DISCUSSÕES

Inicialmente, as constantes de velocidade k e k$_0$ foram obtidas a partir do ajuste de cada isoterma experimental pelos modelos físicos. Esses dados foram organizados na matriz **W$_1$** para serem utilizados no algoritmo da rede. O erro residual de ajuste de cada isoterma pelos modelos individuais para as amostras de ERPU-0%Al$_2$O$_3$ e ERPU-10%Al$_2$O$_3$ são apresentados na Figura 3. Os modelos D3 e Au apresentaram erros muito elevados (9,23 e 5,64, respectivamente) à temperatura de 232°C para a amostra ERPU-0%Al$_2$O$_3$. Já para a amostra da espuma com alumina, os modelos D3 e F1 apresentam erros muito elevados (6,88 e 5,12, respectivamente) à temperatura de 220°C e para efeito de melhor visualização do gráfico, esses dados não estão apresentados na Figura 3.

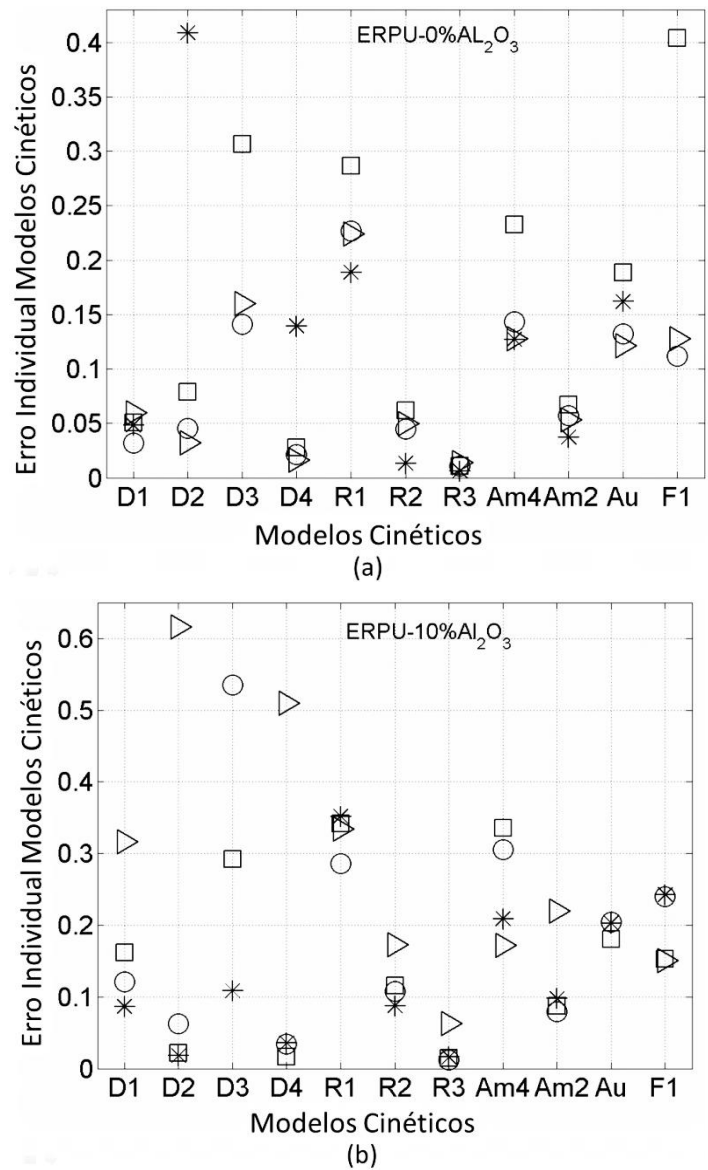

***Figura 3.*** *Erro individual dos modelos cinéticos para as amostras de espumas rígidas de poliuretano contendo glicerol como agente reticulador, a) sem adição de carga contendo alumina (ERPU-0%Al$_2$O$_3$) e b) com adição de 10% de carga contendo alumina (ERPU-10%Al$_2$O$_3$) nas temperaturas (\*)220°C, (O)228°C, (□)230° e (►)232°C, respectivamente*

Como pode ser observado na Figura 03a, em geral, para a amostra sem carga, todos os modelos apresentam elevado erro de ajuste para a isoterma à 232ºC. Por outro lado, os modelos R3, D4 e D2 se mostraram mais adequados, por apresentarem menores erros na descrição das quatro isotermas estudadas. Na Figura 3b, para a amostra com 10% de carga, podemos notar que o modelo cinético de contração volumétrica, R3, é também o modelo cinético que melhor descreve a decomposição térmica do grupo uretana presente na espuma em todas as temperaturas. Para esta amostra, os modelos D4, D2 e R2 também apresentam razoáveis valores de erros de ajuste nas quatro temperaturas.

Uma rede neural MLP foi proposta com o intuito de indicar a contribuição dos modelos físicos que melhor descrevem o processo de decomposição térmica para as duas amostras nas quatro temperaturas. Em uma análise inicial, o estudo foi realizado utilizando-se uma rede neural com todos os modelos físicos sendo considerados como função de ativação na camada intermediária. A Figura 2 apresenta os ajustes promovidos pelas redes neurais às isotermas experimentais das espumas. Como pode ser observado, foi obtida uma excelente concordância entre os dados experimentais de decomposição térmica e o ajuste da rede neural. Os erros residuais obtidos pela rede são da ordem de $10^{-5}$ para ambas as espumas analisadas, conforme apresentado na Tabela 2.

**Tabela 2.** Erro residual da Rede MLP proposta para as espumas rígidas de poliuretano ERPU-0%$Al_2O_3$ e ERPU-10%$Al_2O_3$ em suas respectivas temperaturas

|  | **ERPU** | **Temperatura /°C** | | | |
|---|---|---|---|---|---|
|  |  | **220** | **228** | **230** | **232** |
| Erro Rede / $10^{-5}$ | 0% $Al_2O_3$ | 0,27001 | 1,1780 | 1,0424 | 13,666 |
|  | 10% $Al_2O_3$ | 5,7918 | 1,1453 | 0,74509 | 0,37268 |

A Figura 4 apresenta uma comparação do erro residual da rede MLP e o modelo R3. Como podemos observar, a rede MLP reduziu substancialmente o erro residual em todas as temperaturas para as duas amostras. Esta redução do erro residual da rede frente ao ajuste individual do modelo pode ser explicada com base no fator de correção $W_2$, que introduz um parâmetro de ajuste e indica a contribuição de cada modelo no ajuste total e em casos de modelos assintóticos, como os de Avrami-Erofeev, representa também o valor do resíduo remanescente do processo de decomposição.

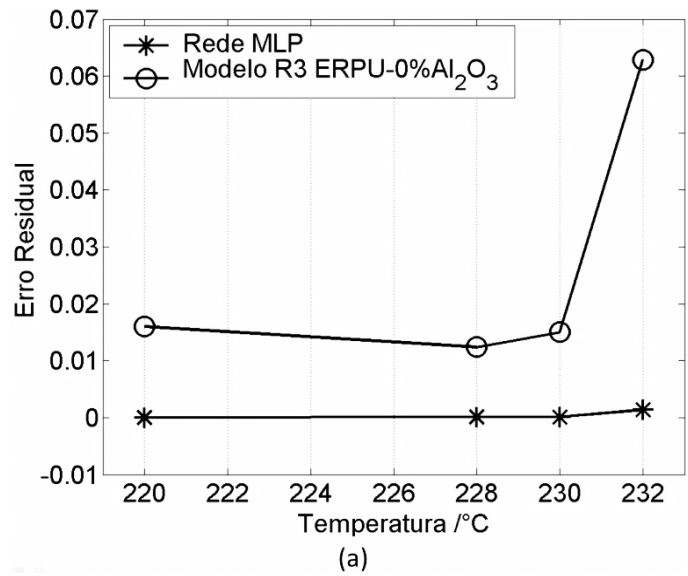

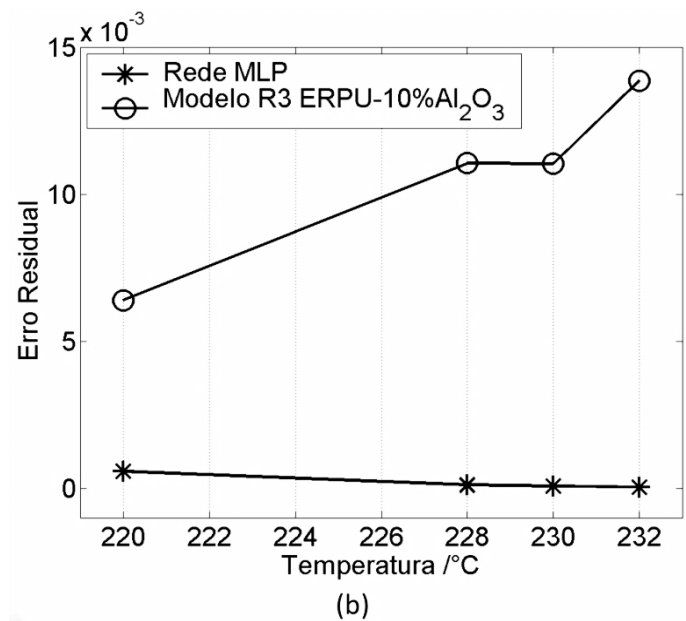

***Figura 4.*** *Erro residual das redes neurais MLP (\*) e do modelo R3(O) para as amostras de espumas rígidas de poliuretano contendo glicerol como agente reticulador, a) sem adição de carga contendo alumina (ERPU-0%Al$_2$O$_3$) e b) com adição de 10% de carga contendo alumina (ERPU-10%Al$_2$O$_3$) nas temperaturas 220°C, 228°C, 230° e 232°C, respectivamente*

Os valores da contribuição de cada modelo, $\mathbf{W_2}$, em cada temperatura estão apresentados na Figura 5. Os valores apresentados estão normalizados e considerados em valor absoluto em cada temperatura.

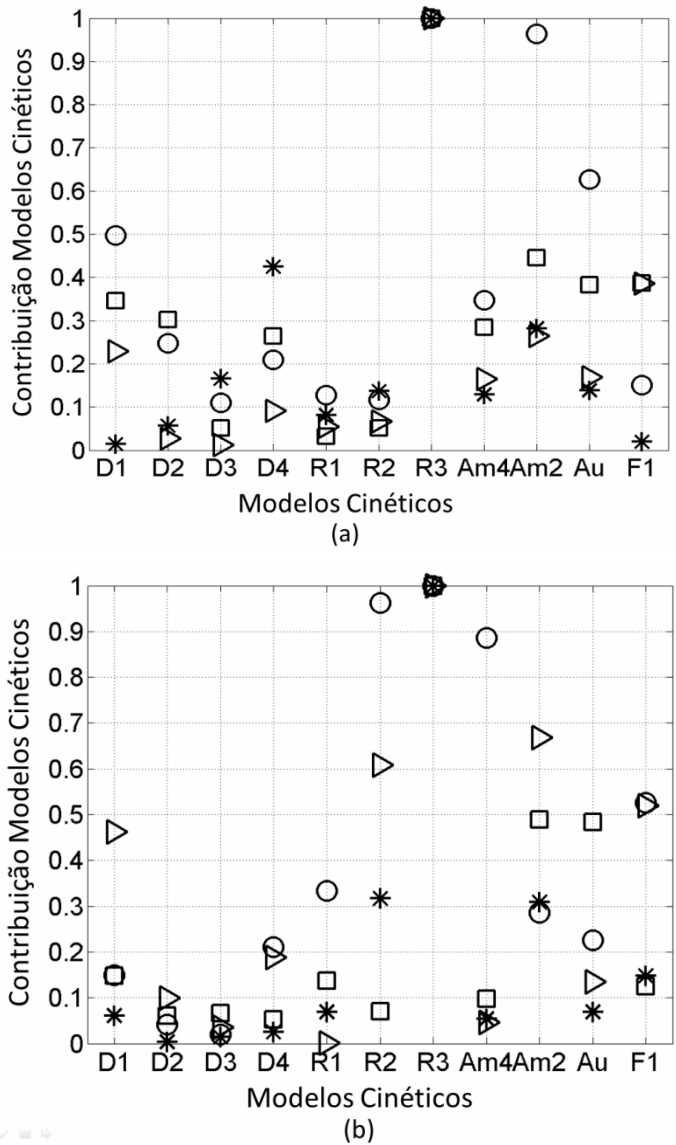

***Figura 5.*** *Contribuição dos modelos cinéticos para as amostras de espumas rígidas de poliuretano contendo glicerol como agente reticulador, a) sem adição de carga contendo alumina (ERPU-0%Al$_2$O$_3$) e b) com adição de 10% de carga contendo alumina (ERPU-10%Al$_2$O$_3$) nas temperaturas (\*)220°C, (O)228°C, (□)230° e (►)232°C, respectivamente, respectivamente*

Analisando a Figura 5 podemos concluir que o modelo cinético de contração volumétrica, R3, é o modelo que mais contribui (e o que possui menor erro individual) para o ajuste da rede MLP em todas as temperaturas analisadas para as duas espumas. Para a espuma sem carga, podemos verificar na Figura 5a, por exemplo, que à temperatura de 228ºC, os quatro modelos cinéticos que mais contribuem no ajuste total da curva são R3, Am2, Au e D1. Embora os modelos que apresentem menor erro de ajuste individual sobre a curva total sejam R3, D4 e D2.

Esta análise é interessante, pois a função matemática que descreve o estado do neurônio da camada de saída é composta por um somatório de onze funções correspondentes aos modelos cinéticos, ponderadas por sua contribuição no ajuste total da curva. Desta forma, nem sempre o modelo que apresenta um erro individual menor é o que tem maior contribuição quando os outros modelos são também considerados, pois cada modelo vai contribuir para determinadas regiões no ajuste total da curva.

Para a amostra com carga à temperatura de 220º C, os modelos que mais contribuem para o ajuste total são R3, R2 e Am2. As curvas experimentais de decomposição das duas amostras diferem pouco, mas conforme explicado anteriormente, a carga de 10% de alumina acrescida nessa espuma de poliuretano não é o suficiente para modificar as propriedades térmicas desse material. Essa carga afeta somente a difusão do oxigênio presente na cadeia polimérica da espuma por meio da formação de uma barreira protetora.[14,16]

Os parâmetros cinéticos de energia de ativação, Ea, e fator de frequência, A, foram calculados considerando-se a teoria de Arrhenius e as constantes de velocidade obtidas por todos os modelos cinéticos nas quatro temperaturas investigadas para as duas amostras. Os resultados são apresentados na Tabela 3. O gráfico de Arrhenius obtido para o modelo R3 é apresentado na Figura 6, em que podemos observar que a linearidade entre os dados garante a validade da teoria.

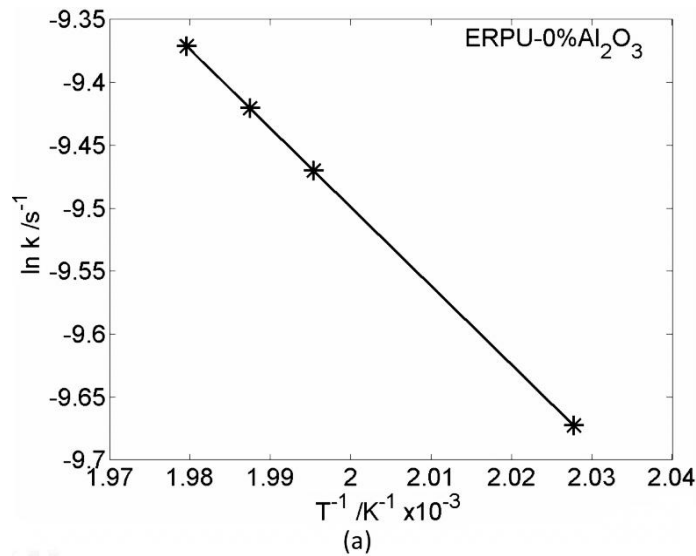

(a)

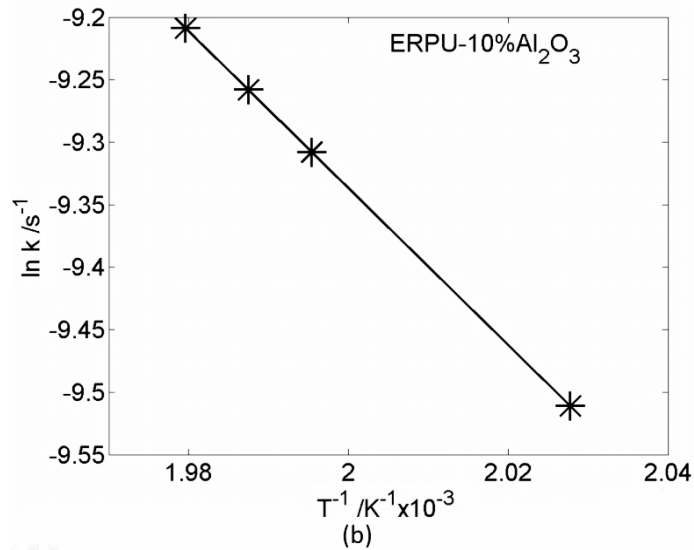

*Figura 6.* Grafico de Arrhenius (ln k x 1/T) para o modelo R3 para as amostras de espumas rígidas de poliuretano contendo glicerol como agente reticulador, a) sem adição de carga contendo alumina (ERPU-0%Al$_2$O$_3$) e b) com adição de 10% de carga contendo alumina (ERPU-10%Al$_2$O$_3$) respectivamente

Considerando o modelo R3 como o modelo que mais contribuiu para o menor erro residual da rede e é também o modelo de menor erro de ajuste individual, o valor da energia de ativação e do fator de frequência para a espuma ERPU-0%Al$_2$O$_3$ é de 56,308 kJ.mol$^{-1}$ e 75,751s$^{-1}$, já para a espuma ERPU-10%Al$_2$O$_3$ os valores são de 65,205 kJ.mol$^{-1}$ e 78,060s$^{-1}$, nessa ordem. Estes valores são condizentes com a literatura.[9,15]

**Tabela 3.** Energia de Ativação e o Fator de Frequência de acordo com os modelos cinéticos para as espumas ERPU-0%Al$_2$O$_3$ e ERPU-10%Al$_2$O$_3$

| Modelo Cinético | ERPU-0%Al$_2$O$_3$ | | ERPU-10%Al$_2$O$_3$ | |
|---|---|---|---|---|
| | Ea /kJ mol$^{-1}$ | A /s$^{-1}$ | Ea /kJ mol$^{-1}$ | A /s$^{-1}$ |
| D$_1$ | 54,038 | 65,785 | 60,052 | 774,77 |
| D$_2$ | 56,295 | 90,030 | 66,161 | 6752,4 |
| D$_3$ | 60,167 | 26,665 | 77,935 | 536,10 |
| D$_4$ | 57,587 | 20,489 | 69,743 | 24,890 |
| R$_1$ | 52,040 | 40,266 | 52,175 | 200,26 |
| R$_2$ | 55,039 | 46,636 | 61,215 | 438,45 |
| R$_3$ | 56,308 | 75,751 | 65,205 | 78,060 |
| Am$_4$ | 57,819 | 122,54 | 57,396 | 734,74 |
| $Am_2$ | 56,905 | 1799,4 | 63,900 | 1176,3 |
| AU  | 62,703 | 479,85 | 60,490 | 1286,4 |
| $F_1$ | 59,049 | 65,785 | 63,755 | 774,77 |

O maior valor de energia de ativação para a espuma com alumina indica que essa necessita de um pouco mais de energia, cerca de 8,897 kJ.mol$^{-1}$, que a espuma sem alumina no processo de decomposição dos grupos uretanos presente em sua estrutura. Fato esse que corrobora com as propriedades retardantes de chama do material, mesmo com a presença de uma pequena quantidade de carga em sua estrutura. Essa barreira energética é responsável pela menor locomoção/difusão do combustível no interior da rede da espuma.

**CONCLUSÃO**

O processo de decomposição térmica de espumas rígidas de poliuretano com glicerol como reticulador, com e sem adição de carga inorgânica, é estudado no presente trabalho. Uma rede neural artificial do tipo MLP foi proposta para o estudo, com a particularidade de assumir os neurônios da camada intermediária como modelos cinéticos de decomposição, uma vez que possuem características semelhantes ao impulso nervoso. A linearização da rede por esta metodologia garante que informações químicas não serão perdidas no processo de ajuste pela rede neural, ou seja, a rede neural, sob esta concepção, não deve ser entendida como um simples mecanismo de ajuste matemático.

Esta rede MLP é uma poderosa ferramenta para tratar o fenômeno de decomposição térmica de sólidos, pois os erros residuais de ajuste são menores do que os resultados obtidos quando são utilizados os modelos individualmente. Além disso, com essa abordagem é possível calcular a contribuição de cada modelo na descrição experimental dos dados. Esta metodologia funciona, pois a rede neural propõe considerar o valor assintótico correto da fração de massa além de alterar a escala de tempo do processo nos modelos cinéticos. Essas alterações permitem a redução no erro residual em média $10^2$ vezes menor, para ambas as espumas, quando comparado com o melhor modelo cinético individual que descreve o processo, R3.

Com este melhor ajuste, a rede neural MLP possibilita o estudo físico detalhado do processo e consequentemente o cálculo mais preciso dos parâmetros cinéticos como a energia de ativação e o fator de frequência. A metodologia apresentada nesse trabalho não é restrita ás espumas rígidas estudas, e pode ser aplicado a outros sistemas, o que sugere um poderoso método de rotina no processo de decomposição térmica de sólidos.